\documentclass{aa}  
\usepackage{graphicx}
\usepackage{txfonts}
\usepackage{txfonts}
\usepackage{xcolor}
\usepackage[colorlinks=true,citecolor=blue,linkcolor=blue]{hyperref}%

\begin{document}


  \title{The jet collimation profile at high resolution in BL Lacertae}

   \author{C.~Casadio
          \inst{1,}\inst{2}
          \and
          N.~R.~MacDonald\inst{2}
          \and
          B.~Boccardi\inst{2}
          \and
          S.~G.~Jorstad\inst{3,}\inst{4} 
          \and
          A.~P.~Marscher\inst{3}
          \and
          T.~P.~Krichbaum\inst{2}
          \and
          J.~A.~Hodgson\inst{5}
          \and
          J-Y.~Kim\inst{2,}\inst{6} 
          \and
          E.~Traianou\inst{2}
          \and
          Z.~R.~Weaver\inst{3}
          \and
          M.~G{\'o}mez~Garrido\inst{7}
          \and
          J.~Gonz{\'a}lez Garc{\'i}a\inst{7}
          \and
          J.~Kallunki\inst{8}
          \and
          M.~Lindqvist\inst{9}
           \and
          S.~S{\'a}nchez\inst{10}
          \and
          J.~Yang\inst{9}
          \and
          J.~A.~Zensus\inst{2}
          }

   \institute{Foundation for Research and Technology - Hellas, IESL \& Institute of Astrophysics, Voutes, 7110 Heraklion, Greece\\
         \email{ccasadio@ia.forth.gr}
         \and
         Max-Planck-Institut f\"ur Radioastronomie, Auf dem   H\"ugel, 69, D-53121 Bonn, Germany
         \and
         Institute for Astrophysical Research, Boston University, Boston, MA 02215, USA 
         \and 
          Astronomical Institute, St. Petersburg State University, St. Petersburg 199034, Russia
          \and
          Department of Physics and Astronomy, Sejong University, 209 Neungdong-ro, Gwangjin-gu, Seoul, South Korea
          \and
          Korea Astronomy and Space Science Institute, Daedeok-daero 776, Yuseong-gu, Daejeon 34055, Republic of Korea
          \and
          Centro de Desarrollos Tecnologicos, Observatorio de Yebes (IGN), 19141 Yebes, Guadalajara, Spain
          \and
          Aalto University Mets\"ahovi Radio Observatory, Mets\"ahovintie 114, 02540 Kylm\"al\"a, Finland
          \and
          Department of Space, Earth and Environment, Chalmers University of Technology, Onsala Space Observatory, 439 92 Onsala, Sweden
          \and
          Instituto de Radio Astronom{\'i}a Milimetrica, Avenida Divina Pastora, 7, Local 20, E-18012 Granada, Spain
            }

\date{Accepted XXX. Received YYY; in original form ZZZ}

 
  \abstract
   {Controversial studies on the jet collimation profile of BL Lacertae (BL Lac), the eponymous blazar 
of BL Lac objects class, complicate the scenario in this already puzzling class of objects. Understanding the jet geometry, in connection with the jet kinematics and the physical conditions in the surrounding medium, is fundamental to better constrain the formation, acceleration and collimation mechanisms in extragalactic jets.}
   {With the aim of investigating the jet geometry in the innermost regions of the jet of BL Lac, and solving the controversy, we explore the radio jet in this source, using high resolution millimeter-wave Very Long Baseline Interferometric (VLBI) data.}
   {We collect 86~GHz GMVA and 43~GHz VLBA data to obtain stacked images that we use to infer the jet collimation profile by means of two comparable methods. We analyze the kinematics at 86 GHz, and we discuss it in the context of the jet expansion. Finally we consider a possible implication of the Bondi sphere in shaping the different expanding region observed along the jet.}
   {The jet in BL Lac expands with an overall conical geometry. A higher expanding rate region is observed between $\sim$ 5 and 10~pc (de-projected) from the black hole. Such a region is associated  with the decrease in brightness usually observed in high-frequency VLBI images of BL Lac. 
   The jet retrieves the original jet expansion around 17~pc, where the presence of a recollimation shock is supported by both the jet profile and the 15~GHz kinematics (MOJAVE survey). The change in the jet expansion profile occurring at $\sim$ 5~pc could be associated with a change in the external pressure profile in correspondence of the Bondi radius ($\sim$3.3$\times$10$^{5}R_{S}$).}
   {}


\titlerunning{The jet collimation profile at high resolution in BL Lac.}

\maketitle



\section{Introduction}

The most accepted scenario to explain the formation of relativistic jets in active galactic nuclei (AGN) involves the extraction of energy from the innermost regions of a black hole's accretion disk \citep{Blandford:1982fr} and/or the ergosphere of the black hole itself \citep{Blandford:1977ys} by large-scale magnetic fields anchored into the magnetosphere of the rotating black hole. This energy, in the form of Poynting flux, is then converted into the kinetic energy of a collimated jet which gradually accelerates until it reaches its asymptotic Lorentz factor where the electromagnetic and kinetic energy fluxes reach equipartition \citep{Vlahakis2004}. In the jet acceleration region, usually denoted as the acceleration and collimation zone (ACZ) \citep[e.g.,][]{Marscher2008}, the jet assumes a parabolic shape (or sometimes even a more collimated cylindrical shape).   

Such a seemingly simple model, however, is under continuous debate, in part due to insufficient observations but also due to the complexity introduced when the mean particle energy is relativistic. How and where the electromagnetic energy is transferred to the jet is an open question. Investigating the transverse jet size and velocity structure at the jet base through VLBI studies is an important tool to study the acceleration mechanisms internal to jets since the acceleration and the geometry of the jet are strictly connected \citep[e.g.,][]{Vlahakis2004, Beskin2006, Lyubarsky2009}.

In particular, mm-VLBI studies (see \cite{Boccardi2017} for a review) provide high resolution images of regions deep in the jet that are otherwise optically thick at longer centimeter wavelengths. Millimeter (mm)-wave VLBI images, therefore, give us a direct probe of the inner geometry and velocity structure of jets in different AGN classes: radio galaxies (Cygnus A - \cite{Boccardi2016a, Boccardi2016b}; M~87 - \cite{Hada2013}, \cite{Kim2018}; NGC~4261 - \cite{Nakahara2018}; NGC~1052 - \cite{Baczko2016,Baczko2019,Nakahara2020}), a narrow-line Seyfert~1 (NLS1) galaxy, 1H~0323+342, \cite{Hada2018},   
and few blazars (Mrk~501 -  \cite{Giroletti2004}; 3C~273 \cite{Akiyama2018}; TXS 2013+370 - \cite{Traianou2020}).  
The picture is further complicated by synchrotron opacity effects \citep{Blandford1979} which can obscure the inner jet and shift the apparent location of the VLBI radio core at large distances from the black hole, $\sim10^{(4-6)}$~Schwarzshild radii ($R_{s}$) \citep[e.g.][]{Marscher2008}, and also by relativistic effects, when the jet is oriented at small viewing angles (as is the case in blazars, i.e., BL Lacs and FSRQs) and the VLBI core emission typically dominates over the fainter extended jet emission. 

Relativistic jets should also be considered in the context of their surrounding environments; the accretion flow, and the surrounding disk winds and interstellar medium. 
The wind (which may be magnetized) coming from the outer parts of the accretion disk may play a crucial role in the confinement of jets \citep{2006MNRAS.368.1561M, 2008MNRAS.388..551T}. \cite{Boccardi2020} show that in radio galaxies of high-excitation (HERG), the ACZ, expressed in Schwarzshild radii, extends over larger distances than it does in low-excitation radio galaxies (LERG) and that this may be associated with the different accretion mechanism in these two classes of objects. HERG, hosting radiatively efficient standard disks, would have a stronger disk wind component that keeps the jet collimated for larger distances.    
This result is in agreement with the behaviour observed in the beamed subset of HERG and LERG, by \cite{Potter2015}. They find that flat spectrum radio quasars (FSRQ), the beamed subgroup of HERG, accelerate over longer distances than BL Lacs, which are instead considered the beamed subgroup of LERG.

Recently \cite{Kovalev2020}, and previously \cite{Pushkarev2017}, investigated the jet geometry of a large sample of AGN using mainly 15 GHz VLBA data from the MOJAVE program. In \cite{Kovalev2020}, they found that ten nearby AGN ($z<0.07$) show a transition from a parabolic to a conical shape in the scales probed at 15 GHz. Among these sources is the jet of BL Lac, whose transition occurs at a linear distance of $\sim$2.5 mas from the black hole ($\sim$24.6~pc de-projected, considering a viewing angle of 7.6$^{\circ}$). In contrast, the earlier study of  \cite{Pushkarev2017}  presented a different result: the jet expands conically from 0.6~mas to 10 mas, thus up $\sim$100~pc de-projected, and no break is observed. In the latter case the jet is not accelerating and has probably ceased to be Poynting flux dominated.     
With the aim of clarifying our understanding of the collimation profile in BL Lac, we perform a detailed study of the jet geometry, now adding also the information from higher frequencies (43 and 86~GHz) VLBI observations. Because of the reduced opacity and the higher resolution ($\sim$ 3 and 6 times higher, for the 43 and 86 GHz observations respectively) our dataset  gains with respect to the 15~GHz data,  we can investigate the innermost portion of the jet in more detail. We demonstrate how higher resolution millimeter VLBI data (with high dynamic range) are essential to determine the jet's collimation profile in the innermost jet regions.

In Section~\ref{obs} we present the data set and the methods used for the analysis and in Section~\ref{jetprofile} we discuss our results, contextualising them within the jet kinematics and the physical conditions of the external medium. Our conclusions are summarized in Section~\ref{sum}.  We adopt a flat cosmology with $\Omega_{m}$= 0.27, $\Omega_{\Lambda}$= 0.73, and $H_{0}$ = 71 km s$^{-1}$ Mpc$^{-1}$ \citep{Komatsu2009}. Assuming these values, at the redshift of BL Lac ($z$=0.0686), 1 mas corresponds to a linear distance of 1.296~pc.

\section{Data and Methods}\label{obs}

We analyze 86~GHz Global mm-VLBI Array (GMVA) data of BL Lac obtained from May 2009 to April 2017 (11 observing epochs) and 43~GHz VLBA data from the VLBA-BU-BLAZAR program\footnote{http://www.bu.edu/blazars/research.html} covering 10 years of observations (94 epochs), from May 2009 to March 2019. 
The GMVA data are part of an on-going monitoring program~\footnote{http://www.bu.edu/blazars/vlbi3mm/}
 (PI: A. Marscher) of a sample of $\gamma$-ray bright blazars and radio galaxies. For the data reduction of both data sets see \cite{Casadio2019} and references therein.   

We used {\tt Difmap} to fit the complex visibilities with a model source consisting of components described by circular Gaussian
brightness distributions. We obtain, for each epoch, a
model-fit that provides information about the full width at half maximum (FWHM) size, the flux density, distance and position angle relative to the core (considered stationary) of each component. For the relative uncertainties we followed \cite{Casadio2019}. Model-fit parameters for all components and epochs at 86 GHz, as used in the analysis, are
reported in Table~\ref{table1}.

\begin{figure}
 \centering
  \includegraphics[width=0.4\textwidth]{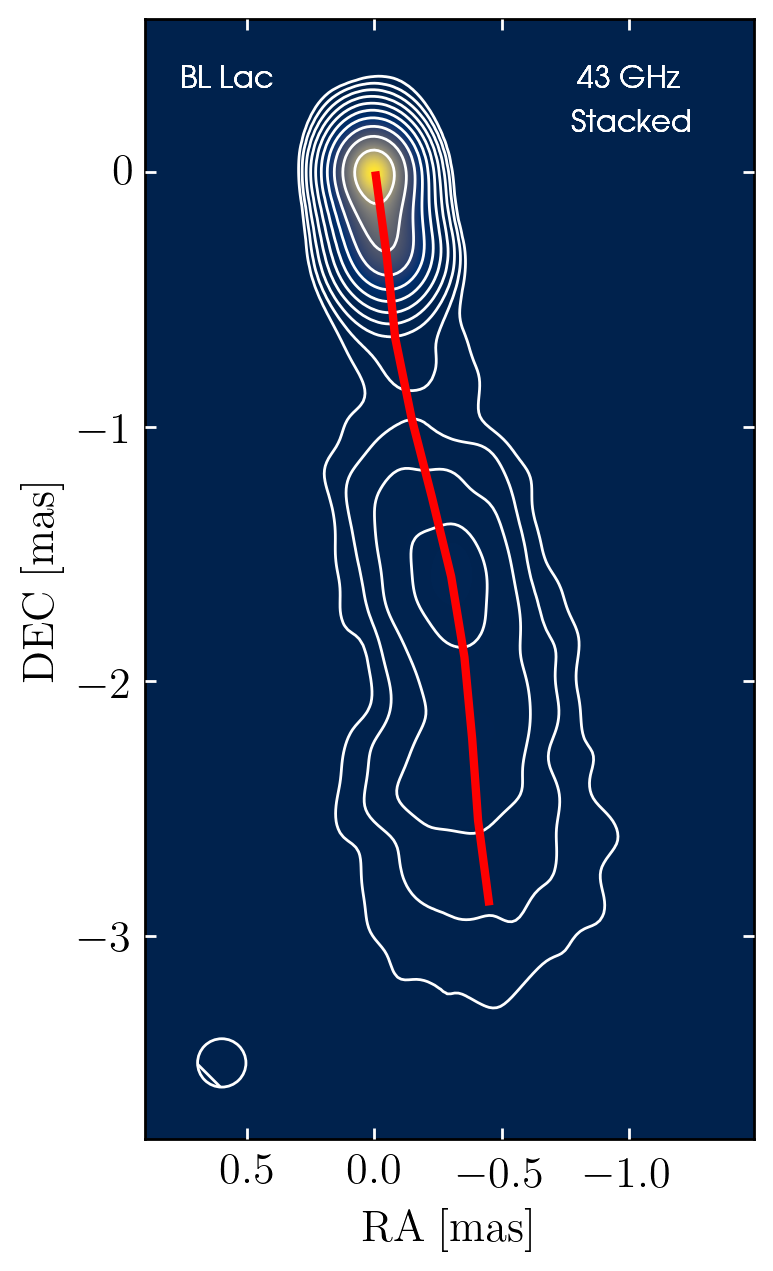}
  \caption{Total intensity stacked image of BL Lac at 43~GHz. 94 images obtained between May 2009 and March 2019 are used. The restoring circular beam \textit{(lower left circle)} is 0.19~mas. Contours are drawn at 0.125, 0.25, 0.5, 1, 2, 4, 8, 16, 32, 64 $\%$ of the peak, 1.71 Jy/beam. The total intensity ridgeline is shown in red.}
  \label{43slice}%
\end{figure}

\begin{figure}
 \centering
  \includegraphics[width=0.4\textwidth]{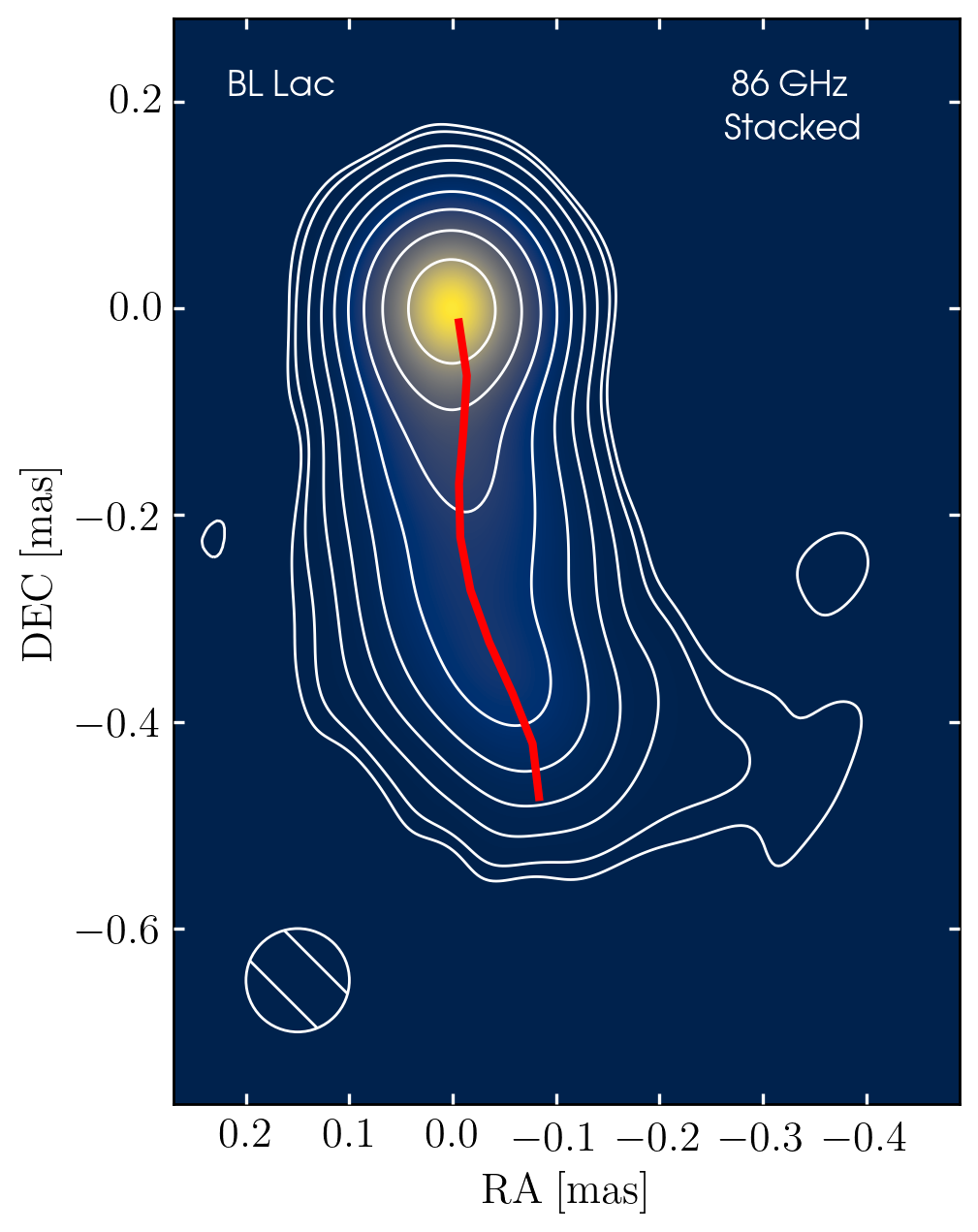}
   \caption{Total intensity stacked image of BL Lac at 86 GHz. 11 images obtained between May 2009 and April 2017 are used. The restoring circular beam \textit{(lower left circle)} is 0.10 mas. Contours are drawn at 0.35, 0.5, 1, 2, 4, 8, 16, 32, 64 $\%$ of the peak, 0.847 Jy/beam. The total intensity ridgeline is shown in red.}
  \label{86slice}%
\end{figure}

 Blazars are highly variable sources and this is mostly reflected in their morphology and behaviour at millimeter wavelengths. Therefore, to obtain a higher dynamic range image and to better recover the entire jet cross-section, we stacked all the total intensity images together.  Before stacking, we convolved the 86 and 43~GHz images with a 0.10~mas and 0.19~mas FWHM restoring circular beams, respectively. The alignment of the images was done using the position of the intensity peak. Our radio map stacking, therefore, implicitly assumes that the radio core in BL~Lac is consistently the brightest feature in the jet at all epochs. While this would not necessarily be true for an FSRQ (e.g., 3C 273) this is indeed generally the case in BL~Lac type objects.  
The resulting stacked images at both frequencies are shown in Figure~\ref{43slice} and~\ref{86slice}, respectively. 

We measured the radial jet width in the stacked images using two different methods: a {\tt Python}~\citep{python} script where the jet ridgeline is obtained first and perpendicular slices are then taken to obtain the transverse jet brightness profile at every cut; and the tasks SLICE and SLFIT in the Astronomical Image Processing System (AIPS; \cite{Greisen1990}) where instead slices have all the same orientation, perpendicular to the main jet direction. The {\tt Python} script (the following packages are included: {\tt matplotlib}\footnote{\cite{matplotlib}}, {\tt numpy}\footnote{\cite{numpy}}, {\tt scipy}\footnote{\cite{scipy}}) should be more sensitive to jet bending, only marginally observed in the considered images. The two methods provide consistent results and the comparison between them was used to test the robustness of the obtained jet profile. Here we will only focus on the first method while we direct the reader to Appendix \ref{app:A} to consult what was obtained using the AIPS procedure. 

We fit every slice brightness profile with a Gaussian distribution and we obtain the deconvolved jet width ($w$) subtracting from the Gaussian FWHM the FWHM of the restoring beam ($b$) in quadrature ($w = \sqrt{FWHM^{2} - b^{2}}$). Following \cite{Pushkarev2017}, we obtain the uncertainties on the jet width measurements rotating the position angle of each slice by $\pm$15$^{\circ}$ in steps of 2$^{\circ}$ and using the new Gaussian profiles to compute the root-mean-square noise.  

In \cite{Marscher2008}, the authors locate the 43~GHz VLBI core at $\sim$0.91~pc (de-projected, assuming 7$^{\circ}$ viewing angle) from the BH, just after the end of the ACZ that terminates at $\sim$0.65~pc from the BH. Given the inverse dependence of the position of the radio core from the jet base with frequency ($r_{core}(\nu)\propto\nu^{-1}$) found in BL Lac, and a core-shift of $\sim$ 0.04~mas between 22 and 43 GHz \citep{O'Sullivan2009}, the core position offset between 43 and 86~GHz is expected to be small. Hence, assuming a negligible core-shift between these two frequencies, we displace the VLBI cores at both frequencies by 0.91~pc and we estimate the radial de-projected distance of each slice from the black hole, using the same viewing angle as in \cite{Marscher2008}.

\section{Results and discussion}
\subsection{The jet collimation profile}
\label{jetprofile}
\begin{figure}
 \centering
  \includegraphics[width=0.52\textwidth]{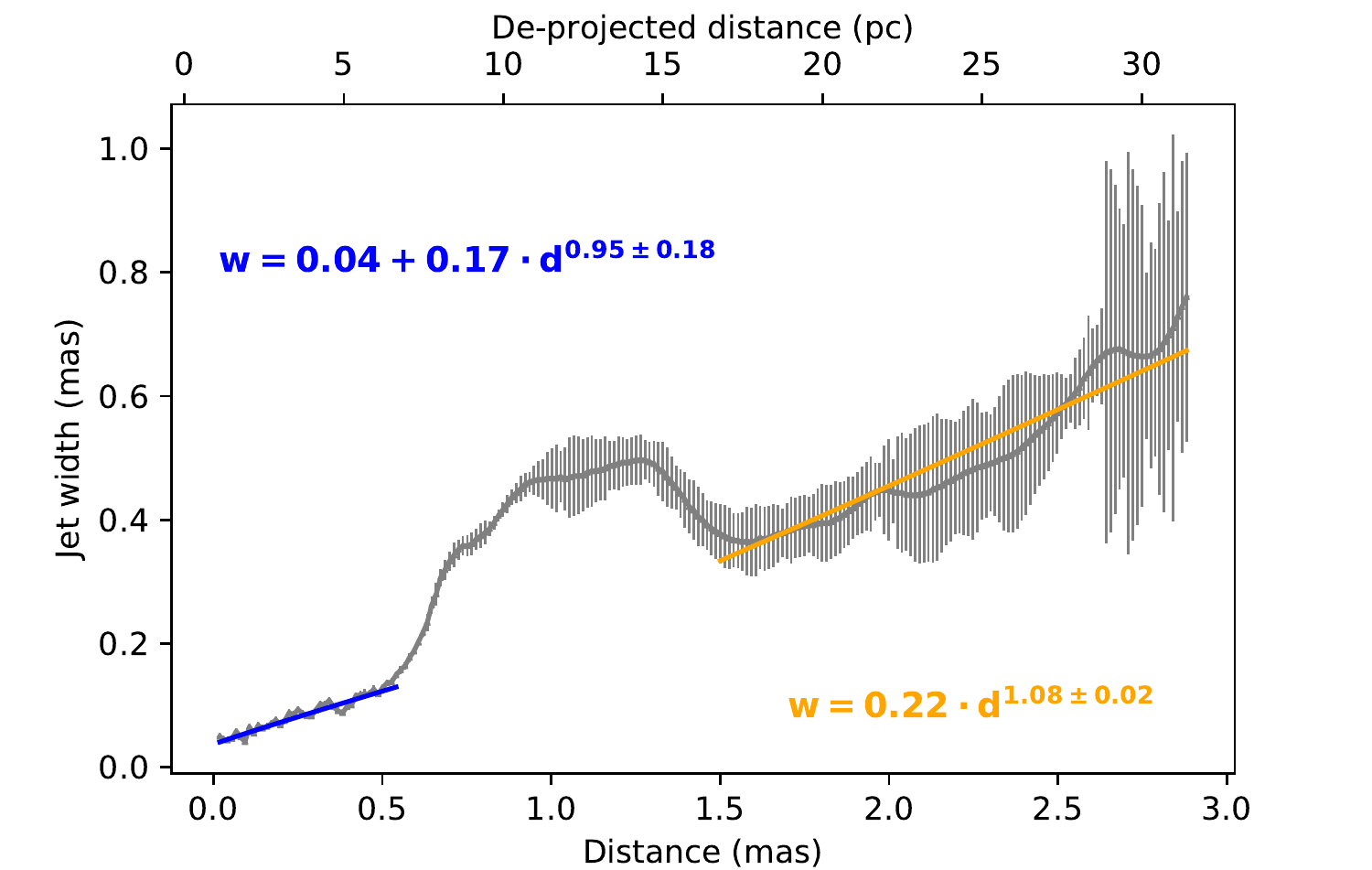}
  \caption{Jet width versus radial distance at 43~GHz, calculated with respect to the VLBI core (in mas) and to the BH (in de-projected parsecs). The blue and orange lines are the best fit of assumed power-laws of the type $w = a + b \cdot d^{k}$ and $w = b \cdot d^{k}$, respectively. The best fit parameters with their uncertainties are also reported in Table~\ref{fit}.}
  \label{43jetprof}%
\end{figure}

\begin{figure}
 \centering
  \includegraphics[width=0.52\textwidth]{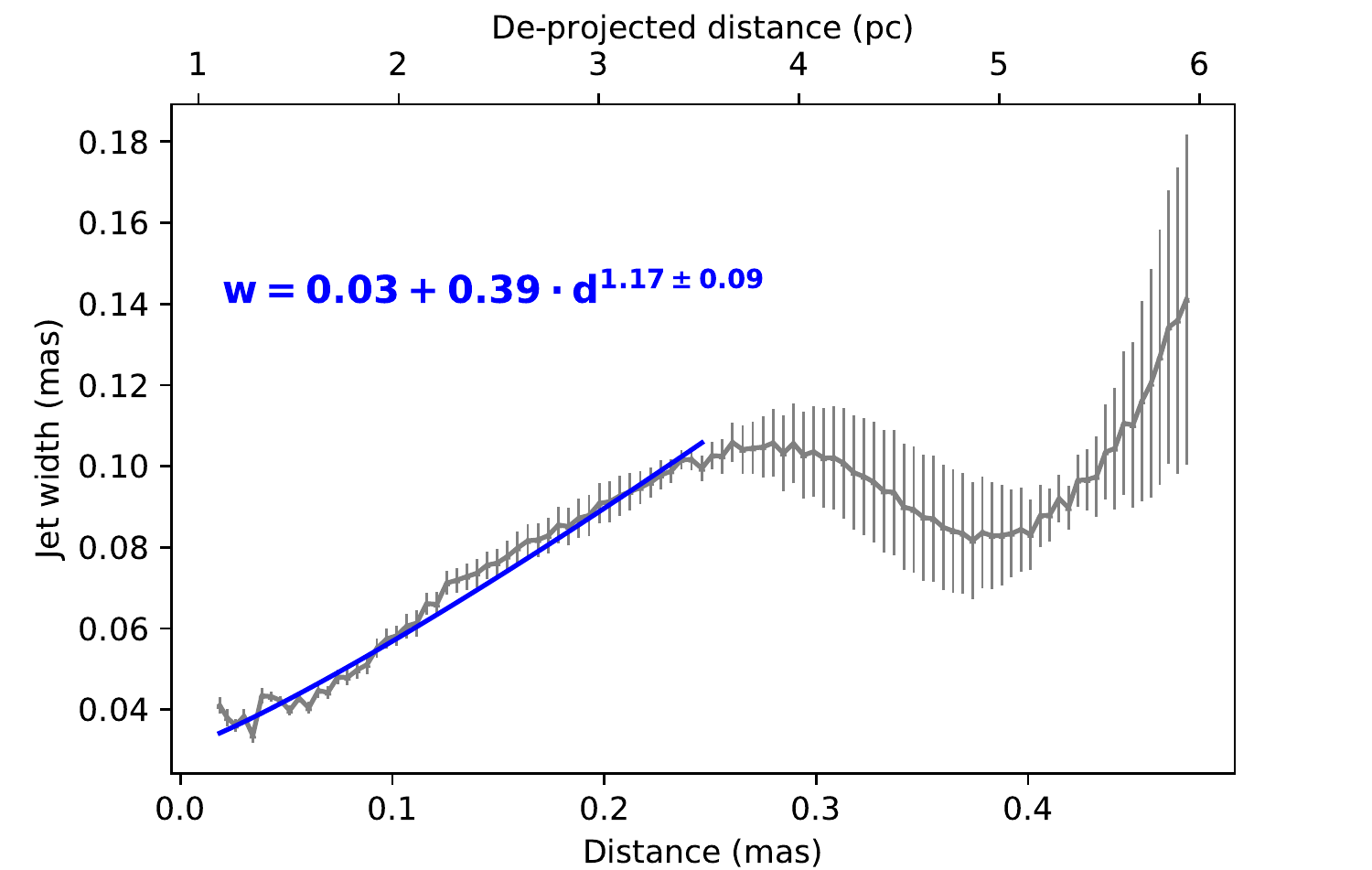}
  \caption{Jet width versus radial distance at 86~GHz, calculated with respect to the VLBI core (in mas) and to the BH (in de-projected parsecs). The blue line is the same as in Figure~\ref{43jetprof}.}
  \label{86jetprof}%
\end{figure}

In Figures~\ref{43jetprof} and~\ref{86jetprof} we analyze the dependence between the jet width $w$ and the de-projected distance from the black hole (in pc), for the 43 and 86~GHz profiles respectively. \textit{It is evident from both profiles that in none of them is it possible to fit the $w-d$ dependence with a single power-law.} The 86~GHz jet expansion profile exhibits a decrease of the jet width around (0.3 -- 0.4)~mas followed by the hint of a more rapid expansion. Indeed, after almost 0.4~mas, as observed at 43~GHz, the jet expands more rapidly and it recollimates later, around 1.5~mas, where it restarts its expansion with a rate similar to the beginning. Therefore, to obtain the information on the overall geometry of the jet, we fit the 86~GHz profile only up to 0.25~mas and we fit separately the part before 0.5~mas and after 1.5~mas at 43~GHz. Using the least squares method, we fit the $w-d$ dependence at 86 GHz and the first 0.5~mas at 43~GHz with a power-law of the type $w = a + b \cdot d^{k}$, where $a$ gives the jet width at the location of the core. We use instead a power-law fit of the type $w = b \cdot d^{k}$ for the 43~GHz jet beyond 1.5~mas. The best fit parameters and relative uncertainties obtained are listed in Table~\ref{fit}. 

The exponent $k$ is expected to be 0.0 for a cylindrical jet, close to 0.5 in case of parabolic expansion and 1.0 instead for a conically expanding jet. We obtain $k\simeq1$ for the three fits, which means that \textit{the overall jet geometry in BL Lac, from $\sim$0.9 to $\sim$30~pc (de-projected), is conical.} 

\begin{table*}[]
    \centering
    \begin{tabular}{c|c|c|c|c}
    Obs. Freq.  & \textit{a} & \textit{b} & \textit{k} &  D \\
    (GHz) & (mas)  & (mas$^{1-k}$) & & (pc)\\
    \hline\hline
    43 & 0.037$\pm$0.007 & 0.166$\pm$0.016 & 0.952$\pm$0.175 & 1.1 - 6.6\\
    & & 0.216$\pm$0.004 & 1.076$\pm$0.024 & 16.8 - 31.4\\
    \hline 
    86 & 0.031$\pm$0.002 & 0.386$\pm$0.044 & 1.167$\pm$ 0.089 & 1.1 - 3.5\\
    \hline
    \end{tabular}
    \caption{Best fit parameters obtained for the power-law equations $w = a + b \cdot d^{k}$ and $w = b \cdot d^{k}$ fitting the jet width - distance dependence at 43 and 86~GHz. In the last column are reported the range of distances from the BH in de-projected parsecs used in each fit.}
    \label{fit}
\end{table*}

\begin{figure*}
 \centering
  \includegraphics[width=0.8\textwidth]{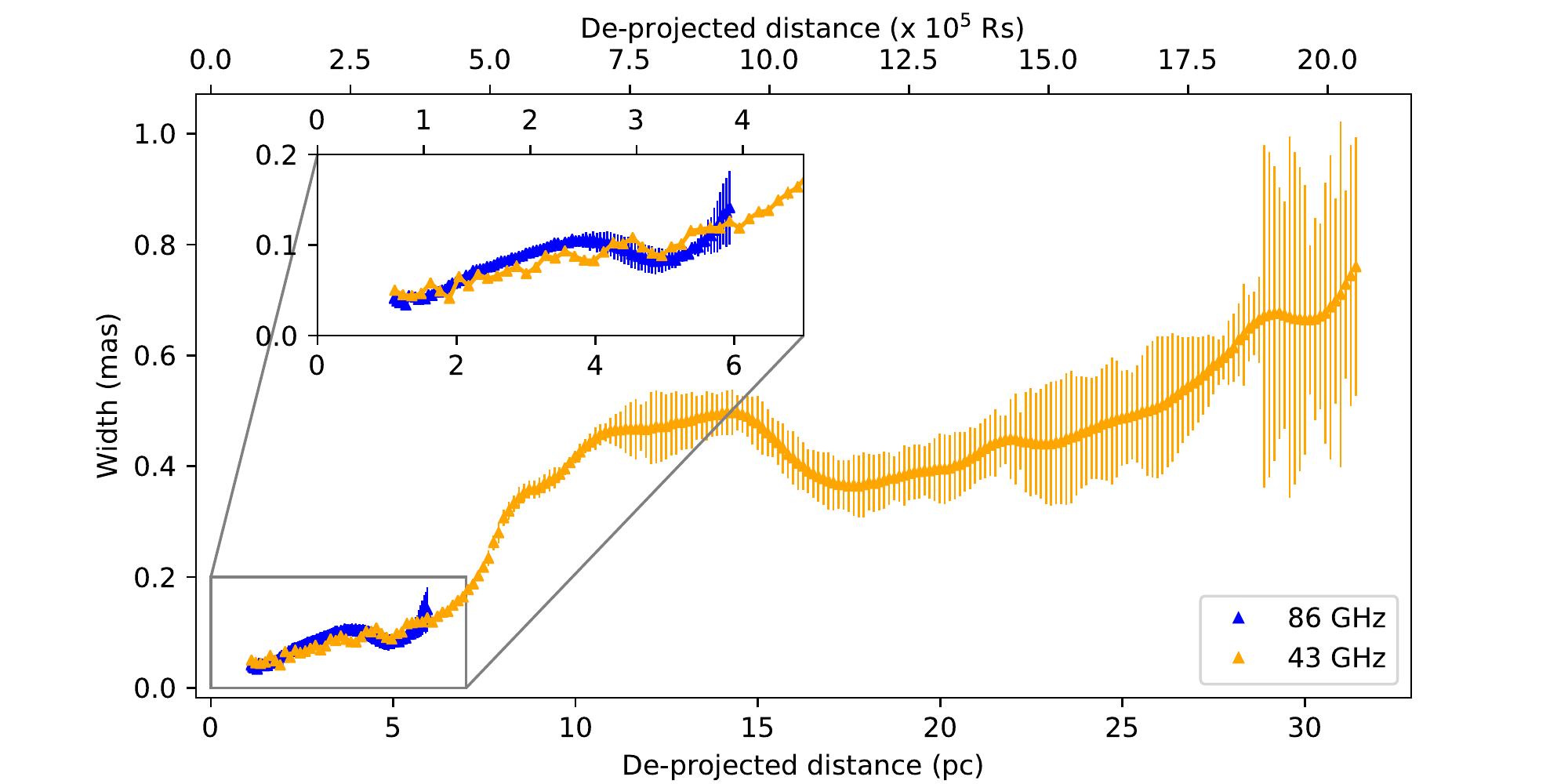}
  \caption{Overlap of the jet collimation profiles in BL Lac as obtained from 43~GHZ VLBA and 86~GHZ GMVA observations. The 86~GHz data define better the starting point of the higher expanding rate region.}
  \label{jetprof}%
\end{figure*}

The derived values for $a$, for the 86 and 43~GHz case, are $a = (31 \pm 2)$ $\mu$as  and $(37 \pm 7)$ $\mu$as. If we compare these values with the core FWHM values obtained from model-fits, we obtain a good agreement within the errors. Indeed, for the 86~GHz core we obtain an average FWHM of 0.023$\pm$0.016~mas, while 0.03$\pm$0.02~mas is the average size of the 43~GHz core as reported in \cite{Jorstad2017}. Given the lower values of the core size at 86~GHz (as inferred from both the fits and the model-fits) than the corresponding ones at 43 GHz, we deduce that in BL Lac the VLBI core at 43 GHz is not completely resolved. To avoid resolution problems, that could artificially flatten the jet collimation profile and mimic a parabolic expansion, \cite{Kovalev2020} consider data starting from 0.9~mas for BL Lac. However, the different expanding rate region observed between $\sim$0.5~mas and 1.5~mas (Figure~\ref{43jetprof}), could introduce a bias in the fit, if it is not excluded from the analysis. Considering data from 0.9~mas, without excluding the aforementioned region, would flatten the jet expansion profile beyond 1.5~mas producing a parabolic geometry instead of a conical expansion. We think this could explain the parabolic shape obtained by \cite{Kovalev2020} in the innermost 2.5~mas of the jet of BL Lac as well as the disagreement between the 15~GHz core size as they obtained fitting the radial width and the visibility data (model-fit components), being the one they obtained from the fit more than two times the one obtained from MOJAVE model-fits.  

 We use the black hole mass as estimated in \cite{Woo2002}, M$_{BH}\sim1.6\times10^{8} M_{\sun}$, to convert the radial de-projected distances into physical distances (Schwarzschild radii, $R_{s}$). The VLBI core, which we assume is co-spatial at the two frequencies, is located at $\sim0.59\times10^{5}R_{S}$ from the black hole. In Figure~\ref{jetprof}, we superimpose the jet collimation profiles in de-projected distances, as obtained at the two frequencies. The higher resolution at 86~GHz helps to constrain better the onset of the higher expanding rate region around 5~pc ($\sim3.3\times10^{5}R_{S}$). Later, at $\sim$ 17~pc ($\sim2.2\times10^{6}R_{S}$) the jet recollimates and returns to expand following the initial expansion profile.

The 3~mm data, stacked together to obtain a higher dynamic range image, are able to recover the same jet width obtained from stacking 7~mm images, but at the same time they enhance the resolution need to shed light on the innermost structure of the jet in BL Lac.

\subsection{The inner jet structure and kinematics}
\begin{figure*}
 \centering
  \includegraphics[width=1.0\textwidth]{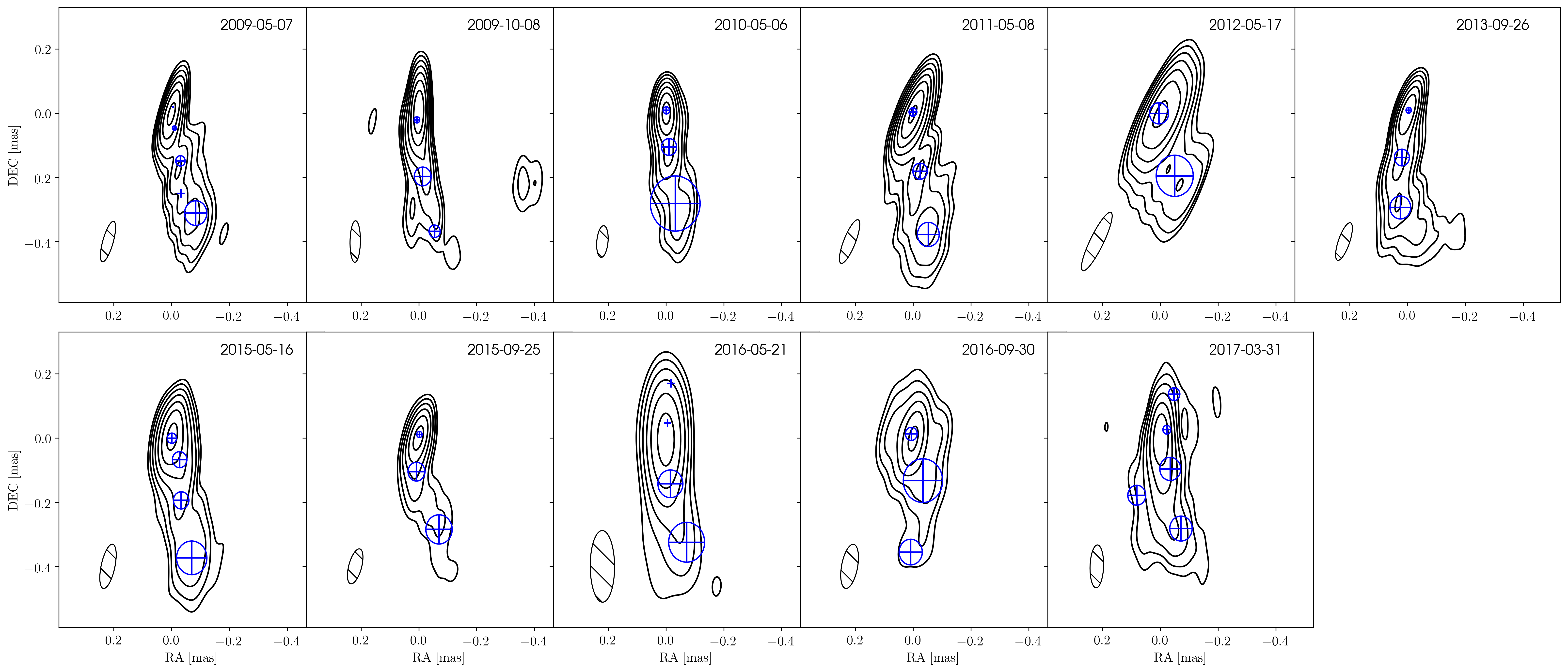}
  \caption{Total intensity images (black contours) of BL Lac obtained with 86 GHz GMVA observations. Contours are drawn at 0.77, 1.53, 3.02, 5.96, 11.74, 23.15, 45.65 and 90 $\%$ of the peak (1.528 Jy/beam) registered in May 2011. The restoring beams, shown at the left, bottom corner, range from 0.04$\times$0.1~mas to 0.08$\times$0.22~mas in FWHM. Model-fits components (blue circles) are overlaid to the maps. In May 2016 and March 2017, an additional component appears upstream the core.}
  \label{multiepoch}%
\end{figure*}

\begin{figure}
 \centering
  \includegraphics[width=0.5\textwidth]{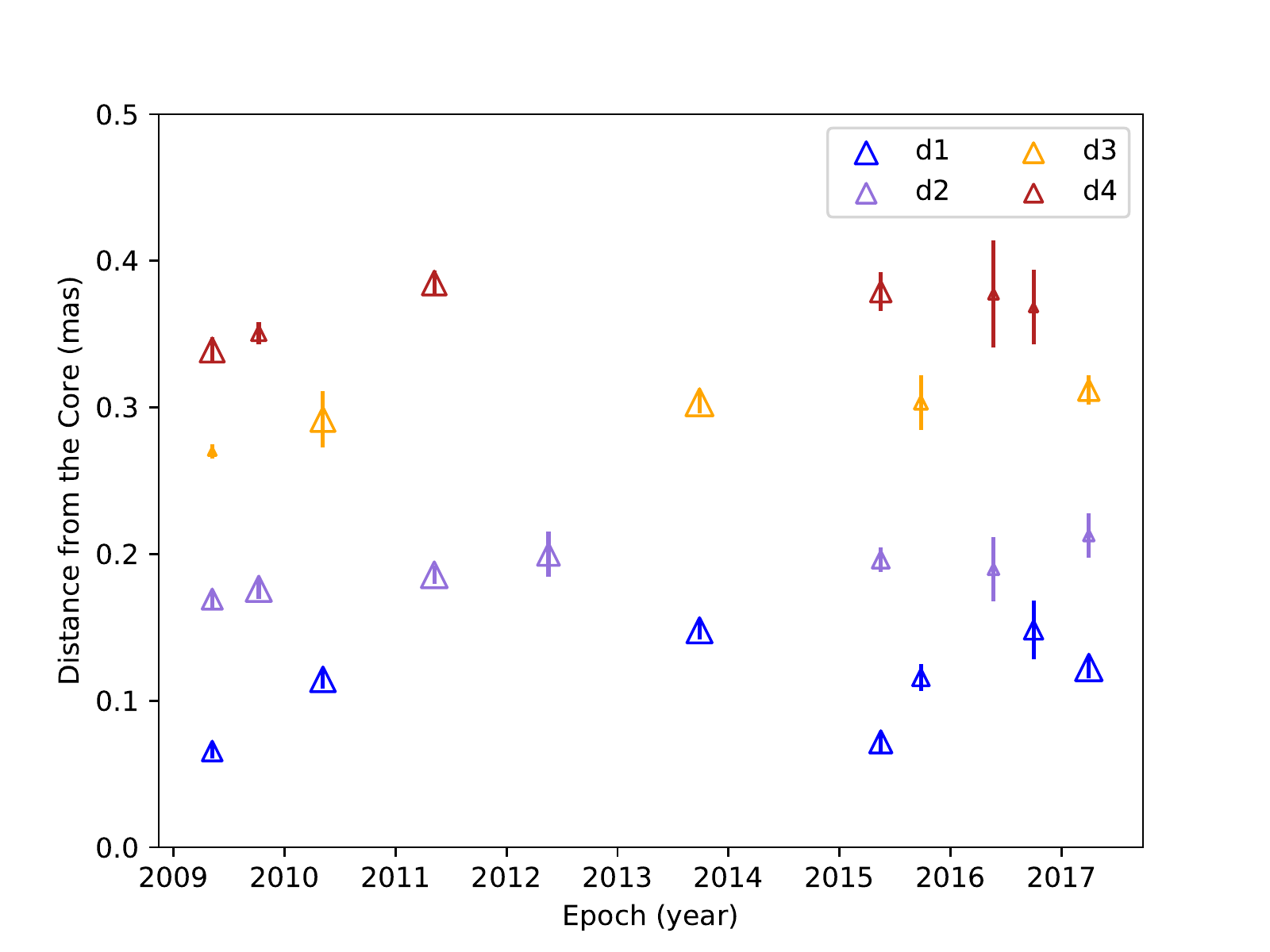}
  \caption{Temporal evolution of the distances from the core of the 86 GHz model-fit components. Symbol sizes are proportional to flux density.}
  \label{kin}%
\end{figure}

In the previous section we reported about the existence of a region in the jet collimation profiles where the jet expands more rapidly, which we excluded from our analysis of the overall jet geometry. We focus here our attention on that region and we try to contextualize it within the jet kinematics and the physical conditions of the external medium.
The kinematics in the first 0.5 mas is rather complex, even when analyzed with the higher resolution provided by the 86 GHz data. Aside from the core, which we consider stationary, we model this first part of the jet with four main stationary components (Figures~\ref{multiepoch} and \ref{kin}). In \cite{Rani2016}, a similar source modelling for BL Lac has been presented for a GMVA data set obtained in May 2013, hence within our observing period. 

\begin{figure}
 \centering
  \includegraphics[width=0.5\textwidth]{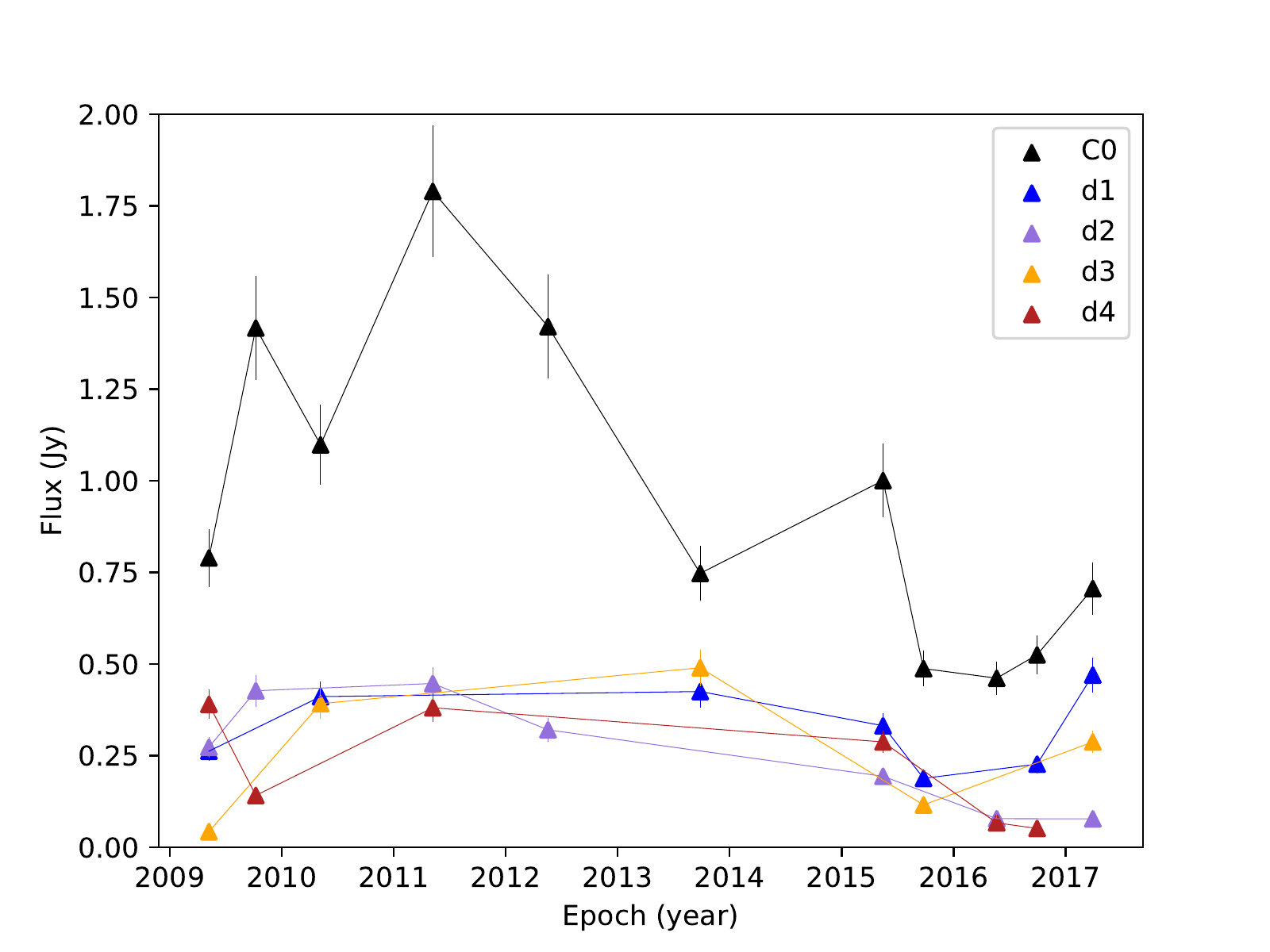}
  \caption{Light curves of the 86 GHz GMVA model-fit components.}
  \label{lc}%
\end{figure}

A quasi-stationary feature, located at 0.26~mas from the 15~GHz VLBI core, has been reported within the MOJAVE survey and interpreted as a recollimation shock \citep{Cohen2014}. If we assume a core-shift of 0.1~mas between the 15~GHz and the 43~GHz core \citep{O'Sullivan2009}, the stationary feature should be located around 0.36~mas in our 86~GHz images. Indeed, the 0.3--0.4~mas region in the 86~GHz jet, is modelled with two components: $d3$, located at $\sim$0.3~mas and $d4$, which shifts in positions between 0.35 and 0.4~mas. The two components are never observed at the same time, aside from the epoch of May 2009 where the identification of $d3$ is doubtful, though. A similar behaviour is observed for the emission around 0.1--0.2~mas. It is hard to say if components $d1-d2$ as well as components $d3-d4$ should be considered separately or if they instead describe the same emitting feature which shifts in positions over time due to core-shift variability. The core-shift variability is strongly related to the flux density variability of the core \citep{Plavin2019} and the 86 GHz core is observed in a rather variable state (Figure~\ref{lc}). The same radio core displacement effect is adopted by \cite{Arshakian2020} to explain the motions of the 15~GHz quasi-stationary feature.

\begin{figure}
 \centering
  \includegraphics[width=0.5\textwidth]{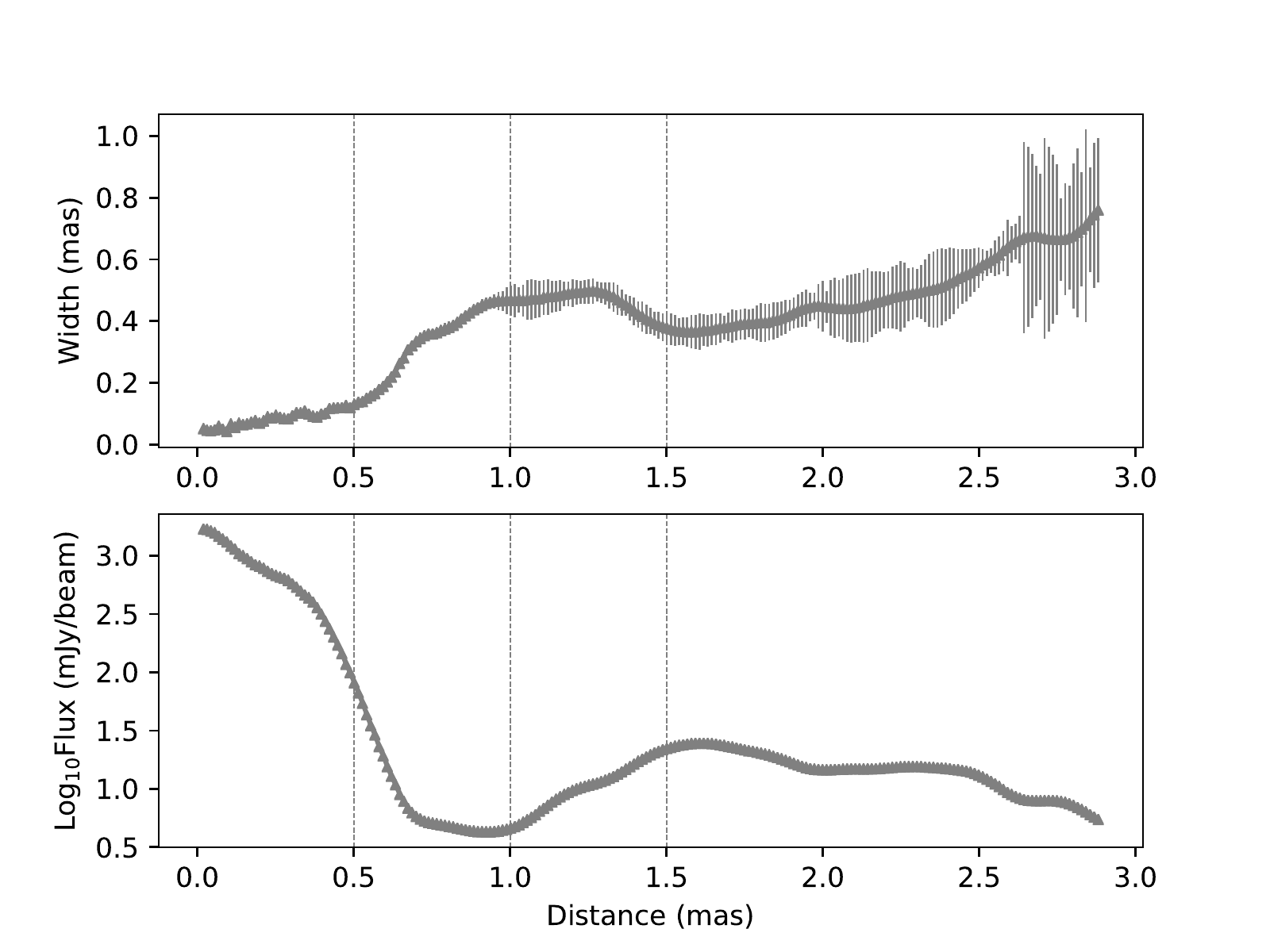}
  \caption{Jet width (top) and flux density (bottom) versus distance along the ridge line at 43 GHz. The jet width and flux density are computed for each transverse slice, as explained in $\S$~\ref{obs}.} 
  \label{jetprofFlux}%
\end{figure}

The location of the stationary emitting region around 0.35~mas coincides with a decrease of the jet width (recollimation), as shown in Figure~\ref{86jetprof}. This would support the nature of the stationary feature as a recollimation shock. 
Moreover, the fact that starting a small distance downstream, the jet expands with a different rate, is hint of change in the physical conditions of either the jet or the external medium (or both). 

While the innermost 0.5~mas of the jet are better explored by the higher resolution provided by the 86~GHz data, the jet geometry and brightness distribution until $\sim$2.9~mas (31~pc, de-projected) are provided by the 43~GHz data-set. As visible from Figure~\ref{jetprofFlux}, the rapid expansion which the jet undergoes from $\sim$0.4~mas to 1~mas is associated with a sharp decrease in total intensity. The flux density goes from $\sim$100~mJy/beam at 0.5~mas to 4.5~mJy/beam at 1 mas and then, it increases again until 1.5~mas where the flux density is back to $\sim$20~mJy/beam. 

\cite{Bach2006} suggest that the low brightness region, visible between $\sim$0.7~mas and 1~mas in 15, 22 and 43~GHz VLBI images, is the result of reduced Doppler boosting associated with either a bending of the jet which becomes misaligned with our line of sight or a change in the jet speed. We instead find that the region of lower brightness usually observed in high-frequency VLBI images of BL Lac is naturally explained by the rapid expansion of the jet and associated adiabatic expansion losses and decrease in magnetic field strength. This is also supported by the fact that the low brightness region has a dependency with frequency, being less pronounced at low frequencies \citep[e.g.,][]{Pearson1988,Bach2006}. When adiabatic losses are the dominant energy loss mechanism, a shift of the turnover frequency toward lower frequencies is indeed expected \citep{Marscher1985}.   

Between 1~mas and 1.5~mas the jet undergoes a compression: the jet width decreases and the flux density increases. The flux density increase could be explained by a local increase in particle density at the site where the jet recollimates or by re-acceleration of relativistic electrons \citep{Mizuno2015,Nishikawa2020}. 

\subsection{Jet expansion and the Bondi radius}

In a conical jet, as is the case for BL Lac, the jet freely expands in the external medium and the magnetic field and electrons Lorentz factor decrease with the expansion. In order to have a change of the \textit{status quo}, a pressure mismatch between the jet and the ambient medium should occur (see for example Fig. 26 in \cite{Fromm2013} for a comparison between a conical and an overpressured jet). A pressure imbalance naturally arises if the ambient pressure decreases and, as a result, a shock wave forms to re-establish the equilibrium between the jet end the external medium \citep[e.g.,][]{Gomez1995,Komissarov1997,Agudo2001}. 

In order to explain the wider jet expansion and further recollimation  observed in BL Lac, we have to assume a change in the external pressure, most probably occurring at the location where the jet changes its expansion rate ($\sim$5 pc, 3.3$\times10^{5}R_{S}$). The Bondi radius, which delimits the sphere of gravitational influence of the black hole and usually extends until a distance of 10$^{5}$-10$^{6}$ R$_{S}$ (\cite{Asada2012}, \cite{Boccardi2020}), could be responsible for this apparent change in the surrounding pressure of the environment into which the jet propagates. Inside the Bondi sphere the external pressure is expected to decrease with distance from the BH. Beyond the Bondi radius instead, the pressure profile becomes flatter, corresponding to the host galaxy's general pressure profile. The jet's inertia causes it to over-expand so that it becomes under-pressured relative to the external medium, causing the condition for a recollimation shock.

The Bondi radius, $r_{B}$, is obtained by setting the escape velocity equal to the sound speed $c_{s}$ ($r_{B}$ = $GMc_{s}^{-2}$). Given the dependence of sound speed on temperature, c$_{s}\sim$10$^{4}T^{1/2}$cm~s$^{-1}$, the Bondi radius can be expressed as follows \citep{Russell2015}: 
\begin{align}
    \frac{r_{B}}{\mathrm{kpc}}= 0.031\left(\frac{kT}{\mathrm{keV}}\right)^{-1}\left(\frac{M_{\mathrm{BH}}}{10^{9}M_{\sun}}\right).
\label{Bondi}
\end{align}
Since we do not have information about the temperature of the accretion flow, but we do have clues regarding the location of the Bondi radius, we can invert the above formula to obtain a range of temperatures, associated with possible locations of the Bondi radius, to compare with values reported in literature for other sources \citep[e.g.][]{Allen2006,Balmaverde2008}. Nevertheless we are aware that the resulting temperatures are sensitive to the assumed black hole mass estimate (M$_{BH}\sim1.6\times10^{8} M_{\sun}$, \citep{Woo2002}).

In the radio galaxy M~87 it has been proposed that a change in the external pressure profile associated with the Bondi sphere is responsible for the jet recollimation occurring at the location of the stationary jet feature HST-1 and for the jet geometry transition (from parabolic to conical) observed at a similar distance \citep{Asada2012}. A possible location for the Bondi radius in BL Lac is the mm-VLBI core, if this one is also associated with a recollimation shock \citep{Gomez2016,Marscher2008} or even upstream, at the end of the ACZ (where the jet should goes from a parabolic to a conical shape), that \cite{Marscher2008} locate 0.65~pc upstream the mm-VLBI core. Therefore, if we place $r_{B}$ at the site of the mm-VLBI core ($\sim$0.9~pc), the electrons temperature would be 5.45~keV, and even higher if we instead consider the end of the ACZ. The obtained temperature, in this case, is too high compared to values found literature. If $r_{B}$ is instead located around 5 pc (where the jet starts expanding faster) and 10~pc (where the jet experiences the maximum width and the minimum flux density) then the electrons temperature in the accretion material would be between 0.99 and 0.33~keV.  In \cite{Balmaverde2008} they infer the temperature at the accretion radius for 44 low power radio galaxies with values ranging between 0.39$\pm$0.10~keV and 1.5$\pm$0.1~keV. Since also BL Lacs are low power sources, considered the beamed subset of low power radio galaxies, it is natural to suppose that the temperature of the accretion flow in BL Lac objects should not deviate much from the values reported in \cite{Balmaverde2008}. We therefore conclude that the Bondi sphere should extend much after the mm-VLBI core and that the change we observe in the jet expansion, around 5~pc de-projected (3.3$\times10^{5}R_{S}$), could be caused by a change in the external pressure profile in proximity of the Bondi radius.  

\subsection{The recollimation shock at $\sim$1.5 mas}
The change in ambient pressure profile that the jet experiences crossing the Bondi sphere causes a pressure mismatch between the jet and the external medium. In these circumstances, numerical hydrodynamical and MHD simulations predict the formation of a recollimation shock \citep[e.g.,][]{Gomez1995,Nishikawa2020} and the acceleration of the bulk flow afterward \citep{Mizuno2015}. The recollimation shock should reduce the Lorentz factor while compressing the plasma. As the plasma flows downstream from the shock, the Lorentz factor increases \citep{Gomez1995, Mizuno2015}. 

We investigated the kinematics at 43~GHz but the identification of components in the region around 1.5~mas is complicated, mainly due to the low brightness region before (discussed previously) which makes it difficult to track the same component in time, as already reported in \cite{Jorstad2017}. Hence we scrutinized the kinematics at 15~GHz, provided by the MOJAVE program, because there the low brightness region is less pronounced. We found that six components (component 4, 5, 6, 11, 16 and 20)\footnote{http://www.physics.purdue.edu/astro/MOJAVE/velocitytableXVIII.html} over the 15 components identified starting from at least 1 mas and moving until at least 2~mas, exhibit a clear acceleration parallel to the velocity vector, after $\sim$1~mas. This may support the presence of a recollimation shock around 1.5~mas, as already suggested by the jet collimation profile (Figure~\ref{43jetprof} and \ref{jetprof}).     

\subsection{Global jet shape}

\begin{figure}
 \centering
  \includegraphics[width=0.5\textwidth]{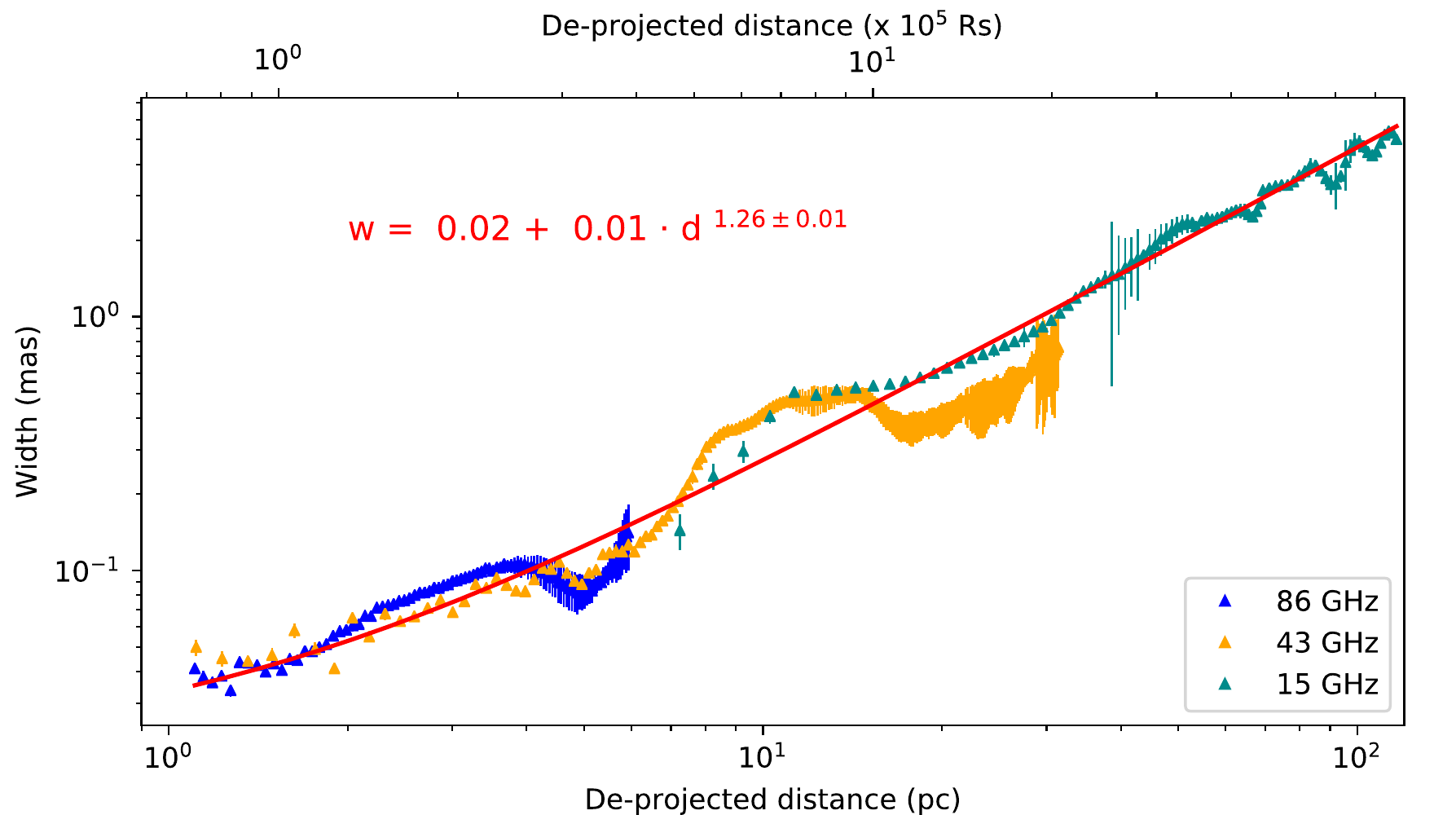}
  \caption{Global jet profile in BL~Lac as obtained from 86 (GMVA), 43 (VLBA-BU-BLAZAR), and 15 GHz (VLBA MOJAVE) data.  The overall jet shape is conical and in agreement at the three frequencies. }
  \label{15G}%
\end{figure}

In Figure~\ref{15G}, we compare the overall jet shape obtained from our 43 and 86 GHz datasets with the one obtained using 15 GHz  MOJAVE data. We downloaded the publicly available stacked image of BL~Lac, which is also the one presented in \cite{Pushkarev2017}, and we obtain the jet collimation profile in the same way we did at the other two frequencies.

We find agreement in the global jet profiles at the three frequencies. The conical shape is well reproduced also by the 15 GHz data, as already found in  \cite{Pushkarev2017}. 
The higher expansion rate region seen at 43 GHz, is also present at 15 GHz, as well as a hint of recollimation, although less pronounced than the one observed at 43 GHz.

\section{Summary}\label{sum}
We studied the jet collimation profile in BL Lac using 3mm GMVA observations, from May 2009 to April 2017, and 7mm VLBA data from the VLBA-BU-BLAZAR program, from May 2009 to March 2019. We built the stacked images at the two frequencies and we analysed the jet width at different separation from the VLBI core. Knowing the BH mass in BL Lac \citep{Woo2002}, the BH - VLBI core distance and viewing angle \citep{Marscher2008}, we also translated the radial distance in de-projected Schwarzschild radii and parsecs from the BH. Our results can be summarized as follows:
\begin{itemize}
    \item The jet expands with an overall conical geometry, which is also confirmed by the analysis of 15 GHz MOJAVE data;
    \item The overall jet expansion is interrupted by a different expanding region between $\sim$ 5~pc and 17~pc, where the jet expands faster and then it recollimates.
    The higher expansion rate is accompanied by a severe decrease in brightness explained considering adiabatic cooling being the main energy loss mechanism there. 
    The jet recollimation, which occurs at around 17~pc downstream, is instead associated with an increase in flux density, explained either by particle re-acceleration or by a local increase in particle density;
    \item The change in the jet expansion profile could be associated with a change in the external pressure profile occurring where the BH ceases its gravitational influence on the surrounding material, at the Bondi radius ($\sim$3.3$\times$10$^{5}R_{S}$);
    \item Both the kinematics and the collimation profile at 3~mm, support the existence of a stationary recollimation shock just before the jet starts expanding faster. The location coincides with the recollimation shock observed at 15~GHz within the MOJAVE survey \citep{Cohen2014,Cohen2015,Arshakian2020}; 
    \item The parallel acceleration displayed by some model-fit components at 15~GHz (MOJAVE survey) after $\sim$1 mas (11 pc), is in agreement with the jet physical conditions downstream a recollimation shock, as predicted by numerical simulations \citep[e.g.,][]{Mizuno2015}. This could prove the existence of a recollimation shock around 17~pc, as already observed in the jet expansion profile and supported by the radial flux density profile. 
     \end{itemize}

In this scenario, the Bondi sphere and so the external medium, play the main role in shaping the jet of BL Lac, which is already freely expanding in the region explored by mm-VLBI data. We think that the jet acceleration region predicted by theoretical models \citep{Vlahakis2004}, ends before the millimeter core, as proposed by \cite{Marscher2008}. There, the main responsible for the jet collimation could be the large scale helical magnetic field observed by \cite{Gomez2016}.  

The case of BL Lac resembles the situation in the radio galaxy M~87, where the stationary feature HST-1 is observed in the proximity of the Bondi radius \citep{Asada2012}. Although in BL Lac we do not observe a change between a parabolic and a conical expansion, as observed in M~87.


\begin{acknowledgements}
We thank Eduardo Ros for the valuable suggestions that helped improve this manuscript. 
This research has been consulting data from the MOJAVE database that is maintained by the MOJAVE team \citep{Lister2009}. C. Casadio acknowledges support from the European Research Council (ERC) under the European Union Horizon 2020 research and innovation programme under the grant agreement No 771282. The research at Boston University was supported by NASA through Fermi Guest Investigator program grants NNX17K0649 an 80NSSC20K1567. This research has made use of data obtained with the Global Millimeter VLBI Array (GMVA), which consists of telescopes operated by the MPIfR, IRAM, Onsala, Metsahovi, Yebes, the Korean VLBI Network, the Greenland
Telescope, the Green Bank Observatory and the Very Long Baseline Array
(VLBA). The VLBA is a facility of the National Science Foundation
operated under cooperative agreement by Associated Universities, Inc.
The data were correlated at the correlator of the MPIfR in Bonn,
Germany. We thank W. Alef, A. Bertarini, H. Rottmann and I. Wagner for their support at the MPIfR VLBI correlator. We also thank Pablo Torne for his help at the IRAM 30m telescope. 

\end{acknowledgements}




\bibliographystyle{aa}

\begin{thebibliography}{56}
\expandafter\ifx\csname natexlab\endcsname\relax\def\natexlab#1{#1}\fi

\bibitem[{Agudo {et~al.}(2001)Agudo, G{\'{o}}mez, Mart{\'{\i}},
  Ib{\'{a}}{\~{n}}ez, Marscher, Alberdi, Aloy, \& Hardee}]{Agudo2001}
Agudo, I., G{\'{o}}mez, J.-L., Mart{\'{\i}}, J.-M., {et~al.} 2001, The
  Astrophysical Journal, 549, L183

\bibitem[{Akiyama {et~al.}(2018)Akiyama, Asada, Fish, Nakamura, Hada, Nagai, \&
  Lonsdale}]{Akiyama2018}
Akiyama, K., Asada, K., Fish, V.~L., {et~al.} 2018, Galaxies, 6

\bibitem[{{Allen} {et~al.}(2006){Allen}, {Dunn}, {Fabian}, {Taylor}, \&
  {Reynolds}}]{Allen2006}
{Allen}, S.~W., {Dunn}, R.~J.~H., {Fabian}, A.~C., {Taylor}, G.~B., \&
  {Reynolds}, C.~S. 2006, \mnras, 372, 21

\bibitem[{{Arshakian} {et~al.}(2020){Arshakian}, {Pushkarev}, {Lister}, \&
  {Savolainen}}]{Arshakian2020}
{Arshakian}, T.~G., {Pushkarev}, A.~B., {Lister}, M.~L., \& {Savolainen}, T.
  2020, \aap, 640, A62

\bibitem[{{Asada} \& {Nakamura}(2012)}]{Asada2012}
{Asada}, K. \& {Nakamura}, M. 2012, \apjl, 745, L28

\bibitem[{{Bach} {et~al.}(2006){Bach}, {Villata}, {Raiteri}, {Agudo}, {Aller},
  {Aller}, {Denn}, {G{\'o}mez}, {Jorstad}, {Marscher}, {Mutel}, \&
  {Ter{\"a}sranta}}]{Bach2006}
{Bach}, U., {Villata}, M., {Raiteri}, C.~M., {et~al.} 2006, \aap, 456, 105

\bibitem[{{Baczko} {et~al.}(2019){Baczko}, {Schulz}, {Kadler}, {Ros},
  {Perucho}, {Fromm}, \& {Wilms}}]{Baczko2019}
{Baczko}, A.~K., {Schulz}, R., {Kadler}, M., {et~al.} 2019, \aap, 623, A27

\bibitem[{{Baczko} {et~al.}(2016){Baczko}, {Schulz}, {Kadler}, {Ros},
  {Perucho}, {Krichbaum}, {B{\"o}ck}, {Bremer}, {Grossberger}, {Lindqvist},
  {Lobanov}, {Mannheim}, {Mart{\'\i}-Vidal}, {M{\"u}ller}, {Wilms}, \&
  {Zensus}}]{Baczko2016}
{Baczko}, A.~K., {Schulz}, R., {Kadler}, M., {et~al.} 2016, \aap, 593, A47

\bibitem[{{Balmaverde} {et~al.}(2008){Balmaverde}, {Baldi}, \&
  {Capetti}}]{Balmaverde2008}
{Balmaverde}, B., {Baldi}, R.~D., \& {Capetti}, A. 2008, \aap, 486, 119

\bibitem[{{Beskin} \& {Nokhrina}(2006)}]{Beskin2006}
{Beskin}, V.~S. \& {Nokhrina}, E.~E. 2006, \mnras, 367, 375

\bibitem[{{Blandford} \& {K{\"o}nigl}(1979)}]{Blandford1979}
{Blandford}, R.~D. \& {K{\"o}nigl}, A. 1979, \apj, 232, 34

\bibitem[{Blandford \& Payne(1982)}]{Blandford:1982fr}
Blandford, R.~D. \& Payne, D.~G. 1982, Monthly Notices of the Royal
  Astronomical Society, 199, 883

\bibitem[{Blandford \& Znajek(1977)}]{Blandford:1977ys}
Blandford, R.~D. \& Znajek, R.~L. 1977, Monthly Notices of the Royal
  Astronomical Society, 179, 433

\bibitem[{{Boccardi} {et~al.}(2016{\natexlab{a}}){Boccardi}, {Krichbaum},
  {Bach}, {Bremer}, \& {Zensus}}]{Boccardi2016a}
{Boccardi}, B., {Krichbaum}, T.~P., {Bach}, U., {Bremer}, M., \& {Zensus},
  J.~A. 2016{\natexlab{a}}, \aap, 588, L9

\bibitem[{{Boccardi} {et~al.}(2016{\natexlab{b}}){Boccardi}, {Krichbaum},
  {Bach}, {Mertens}, {Ros}, {Alef}, \& {Zensus}}]{Boccardi2016b}
{Boccardi}, B., {Krichbaum}, T.~P., {Bach}, U., {et~al.} 2016{\natexlab{b}},
  \aap, 585, A33

\bibitem[{{Boccardi} {et~al.}(2017){Boccardi}, {Krichbaum}, {Ros}, \&
  {Zensus}}]{Boccardi2017}
{Boccardi}, B., {Krichbaum}, T.~P., {Ros}, E., \& {Zensus}, J.~A. 2017, \aapr,
  25, 4

\bibitem[{{Boccardi} {et~al.}(2020){Boccardi}, {Perucho}, {Casadio}, {Grandi},
  {Macconi}, {Torresi}, {Pellegrini}, {Krichbaum}, {Kadler}, {Giovannini},
  {Karamanavis}, {Ricci}, {Madika}, {Bach}, {Ros}, {Giroletti}, \&
  {Zensus}}]{Boccardi2020}
{Boccardi}, B., {Perucho}, M., {Casadio}, C., {et~al.} 2020, arXiv e-prints,
  arXiv:2012.14831

\bibitem[{{Casadio} {et~al.}(2019){Casadio}, {Marscher}, {Jorstad}, {Blinov},
  {MacDonald}, {Krichbaum}, {Boccardi}, {Traianou}, {G{\'o}mez}, {Agudo},
  {Sohn}, {Bremer}, {Hodgson}, {Kallunki}, {Kim}, {Williamson}, \&
  {Zensus}}]{Casadio2019}
{Casadio}, C., {Marscher}, A.~P., {Jorstad}, S.~G., {et~al.} 2019, \aap, 622,
  A158

\bibitem[{{Cohen} {et~al.}(2015){Cohen}, {Meier}, {Arshakian}, {Clausen-Brown},
  {Homan}, {Hovatta}, {Kovalev}, {Lister}, {Pushkarev}, {Richards}, \&
  {Savolainen}}]{Cohen2015}
{Cohen}, M.~H., {Meier}, D.~L., {Arshakian}, T.~G., {et~al.} 2015, \apj, 803, 3

\bibitem[{{Cohen} {et~al.}(2014){Cohen}, {Meier}, {Arshakian}, {Homan},
  {Hovatta}, {Kovalev}, {Lister}, {Pushkarev}, {Richards}, \&
  {Savolainen}}]{Cohen2014}
{Cohen}, M.~H., {Meier}, D.~L., {Arshakian}, T.~G., {et~al.} 2014, \apj, 787,
  151

\bibitem[{{Fromm} {et~al.}(2013){Fromm}, {Ros}, {Perucho}, {Savolainen},
  {Mimica}, {Kadler}, {Lobanov}, \& {Zensus}}]{Fromm2013}
{Fromm}, C.~M., {Ros}, E., {Perucho}, M., {et~al.} 2013, \aap, 557, A105

\bibitem[{{Giroletti} {et~al.}(2004){Giroletti}, {Giovannini}, {Feretti},
  {Cotton}, {Edwards}, {Lara}, {Marscher}, {Mattox}, {Piner}, \&
  {Venturi}}]{Giroletti2004}
{Giroletti}, M., {Giovannini}, G., {Feretti}, L., {et~al.} 2004, \apj, 600, 127

\bibitem[{{G{\'o}mez} {et~al.}(2016){G{\'o}mez}, {Lobanov}, {Bruni}, {Kovalev},
  {Marscher}, {Jorstad}, {Mizuno}, {Bach}, {Sokolovsky}, {Anderson}, {Galindo},
  {Kardashev}, \& {Lisakov}}]{Gomez2016}
{G{\'o}mez}, J.~L., {Lobanov}, A.~P., {Bruni}, G., {et~al.} 2016, \apj, 817, 96

\bibitem[{{G{\'o}mez} {et~al.}(1995){G{\'o}mez}, {Marti}, {Marscher}, {Ibanez},
  \& {Marcaide}}]{Gomez1995}
{G{\'o}mez}, J.~L., {Marti}, J.~M.~A., {Marscher}, A.~P., {Ibanez}, J.~M.~A.,
  \& {Marcaide}, J.~M. 1995, \apjl, 449, L19

\bibitem[{{Greisen}(1990)}]{Greisen1990}
{Greisen}, E.~W. 1990, in Acquisition, Processing and Archiving of Astronomical
  Images, 125--142

\bibitem[{{Hada} {et~al.}(2018){Hada}, {Doi}, {Wajima}, {D'Ammand o},
  {Orienti}, {Giroletti}, {Giovannini}, {Nakamura}, \& {Asada}}]{Hada2018}
{Hada}, K., {Doi}, A., {Wajima}, K., {et~al.} 2018, \apj, 860, 141

\bibitem[{{Hada} {et~al.}(2013){Hada}, {Kino}, {Doi}, {Nagai}, {Honma},
  {Hagiwara}, {Giroletti}, {Giovannini}, \& {Kawaguchi}}]{Hada2013}
{Hada}, K., {Kino}, M., {Doi}, A., {et~al.} 2013, \apj, 775, 70

\bibitem[{Hunter(2007)}]{matplotlib}
Hunter, J.~D. 2007, Computing in Science \& Engineering, 9, 90

\bibitem[{{Jorstad} {et~al.}(2017){Jorstad}, {Marscher}, {Morozova},
  {Troitsky}, {Agudo}, {Casadio}, {Foord}, {G{\'o}mez}, {MacDonald}, {Molina},
  {L{\"a}hteenm{\"a}ki}, {Tammi}, \& {Tornikoski}}]{Jorstad2017}
{Jorstad}, S.~G., {Marscher}, A.~P., {Morozova}, D.~A., {et~al.} 2017, \apj,
  846, 98

\bibitem[{{Kim} {et~al.}(2018){Kim}, {Krichbaum}, {Lu}, {Ros}, {Bach},
  {Bremer}, {de Vicente}, {Lindqvist}, \& {Zensus}}]{Kim2018}
{Kim}, J.~Y., {Krichbaum}, T.~P., {Lu}, R.~S., {et~al.} 2018, \aap, 616, A188

\bibitem[{{Komatsu} {et~al.}(2009){Komatsu}, {Dunkley}, {Nolta}, {Bennett},
  {Gold}, {Hinshaw}, {Jarosik}, {Larson}, {Limon}, {Page}, {Spergel},
  {Halpern}, {Hill}, {Kogut}, {Meyer}, {Tucker}, {Weiland}, {Wollack}, \&
  {Wright}}]{Komatsu2009}
{Komatsu}, E., {Dunkley}, J., {Nolta}, M.~R., {et~al.} 2009, \apjs, 180, 330

\bibitem[{Komissarov \& Falle(1997)}]{Komissarov1997}
Komissarov, S.~S. \& Falle, S. A. E.~G. 1997, Monthly Notices of the Royal
  Astronomical Society, 288, 833

\bibitem[{{Kovalev} {et~al.}(2020){Kovalev}, {Pushkarev}, {Nokhrina}, {Plavin},
  {Beskin}, {Chernoglazov}, {Lister}, \& {Savolainen}}]{Kovalev2020}
{Kovalev}, Y.~Y., {Pushkarev}, A.~B., {Nokhrina}, E.~E., {et~al.} 2020, \mnras,
  495, 3576

\bibitem[{{Lister} {et~al.}(2009){Lister}, {Cohen}, {Homan}, {Kadler},
  {Kellermann}, {Kovalev}, {Ros}, {Savolainen}, \& {Zensus}}]{Lister2009}
{Lister}, M.~L., {Cohen}, M.~H., {Homan}, D.~C., {et~al.} 2009, \aj, 138, 1874

\bibitem[{{Lyubarsky}(2009)}]{Lyubarsky2009}
{Lyubarsky}, Y. 2009, \apj, 698, 1570

\bibitem[{{Marscher} \& {Gear}(1985)}]{Marscher1985}
{Marscher}, A.~P. \& {Gear}, W.~K. 1985, \apj, 298, 114

\bibitem[{{Marscher} {et~al.}(2008){Marscher}, {Jorstad}, {D'Arcangelo},
  {Smith}, {Williams}, {Larionov}, {Oh}, {Olmstead}, {Aller}, {Aller},
  {McHardy}, {L{\"a}hteenm{\"a}ki}, {Tornikoski}, {Valtaoja}, {Hagen-Thorn},
  {Kopatskaya}, {Gear}, {Tosti}, {Kurtanidze}, {Nikolashvili}, {Sigua},
  {Miller}, \& {Ryle}}]{Marscher2008}
{Marscher}, A.~P., {Jorstad}, S.~G., {D'Arcangelo}, F.~D., {et~al.} 2008, \nat,
  452, 966

\bibitem[{{McKinney}(2006)}]{2006MNRAS.368.1561M}
{McKinney}, J.~C. 2006, \mnras, 368, 1561

\bibitem[{{Mizuno} {et~al.}(2015){Mizuno}, {G{\'o}mez}, {Nishikawa}, {Meli},
  {Hardee}, \& {Rezzolla}}]{Mizuno2015}
{Mizuno}, Y., {G{\'o}mez}, J.~L., {Nishikawa}, K.-I., {et~al.} 2015, \apj, 809,
  38

\bibitem[{{Nakahara} {et~al.}(2018){Nakahara}, {Doi}, {Murata}, {Hada},
  {Nakamura}, \& {Asada}}]{Nakahara2018}
{Nakahara}, S., {Doi}, A., {Murata}, Y., {et~al.} 2018, \apj, 854, 148

\bibitem[{{Nakahara} {et~al.}(2020){Nakahara}, {Doi}, {Murata}, {Nakamura},
  {Hada}, {Asada}, {Sawada-Satoh}, \& {Kameno}}]{Nakahara2020}
{Nakahara}, S., {Doi}, A., {Murata}, Y., {et~al.} 2020, \aj, 159, 14

\bibitem[{{Nishikawa} {et~al.}(2020){Nishikawa}, {Mizuno}, {G{\'o}mez},
  {Du{\c{t}}an}, {Niemiec}, {Kobzar}, {MacDonald}, {Meli}, {Pohl}, \&
  {Hirotani}}]{Nishikawa2020}
{Nishikawa}, K., {Mizuno}, Y., {G{\'o}mez}, J.~L., {et~al.} 2020, \mnras, 493,
  2652

\bibitem[{{O'Sullivan} \& {Gabuzda}(2009)}]{O'Sullivan2009}
{O'Sullivan}, S.~P. \& {Gabuzda}, D.~C. 2009, \mnras, 400, 26

\bibitem[{{Pearson} \& {Readhead}(1988)}]{Pearson1988}
{Pearson}, T.~J. \& {Readhead}, A.~C.~S. 1988, \apj, 328, 114

\bibitem[{{Plavin} {et~al.}(2019){Plavin}, {Kovalev}, {Pushkarev}, \&
  {Lobanov}}]{Plavin2019}
{Plavin}, A.~V., {Kovalev}, Y.~Y., {Pushkarev}, A.~B., \& {Lobanov}, A.~P.
  2019, \mnras, 485, 1822

\bibitem[{{Potter} \& {Cotter}(2015)}]{Potter2015}
{Potter}, W.~J. \& {Cotter}, G. 2015, \mnras, 453, 4070

\bibitem[{{Pushkarev} {et~al.}(2017){Pushkarev}, {Kovalev}, {Lister}, \&
  {Savolainen}}]{Pushkarev2017}
{Pushkarev}, A.~B., {Kovalev}, Y.~Y., {Lister}, M.~L., \& {Savolainen}, T.
  2017, \mnras, 468, 4992

\bibitem[{{Rani} {et~al.}(2016){Rani}, {Krichbaum}, {Hodgson}, {Koyama},
  {Zensus}, {Fuhramnn}, {Marscher}, \& {Jorstad}}]{Rani2016}
{Rani}, B., {Krichbaum}, T., {Hodgson}, J., {et~al.} 2016, Galaxies, 4, 32

\bibitem[{{Russell} {et~al.}(2015){Russell}, {Fabian}, {McNamara}, \&
  {Broderick}}]{Russell2015}
{Russell}, H.~R., {Fabian}, A.~C., {McNamara}, B.~R., \& {Broderick}, A.~E.
  2015, \mnras, 451, 588

\bibitem[{{Tchekhovskoy} {et~al.}(2008){Tchekhovskoy}, {McKinney}, \&
  {Narayan}}]{2008MNRAS.388..551T}
{Tchekhovskoy}, A., {McKinney}, J.~C., \& {Narayan}, R. 2008, \mnras, 388, 551

\bibitem[{{Traianou} {et~al.}(2020){Traianou}, {Krichbaum}, {Boccardi},
  {Angioni}, {Rani}, {Liu}, {Ros}, {Bach}, {Sokolovsky}, {Lisakov},
  {Kiehlmann}, {Gurwell}, \& {Zensus}}]{Traianou2020}
{Traianou}, E., {Krichbaum}, T.~P., {Boccardi}, B., {et~al.} 2020, \aap, 634,
  A112

\bibitem[{{Van der Walt} {et~al.}(2011){Van der Walt}, {Colbert}, \&
  {Varoquaux}}]{numpy}
{Van der Walt}, S., {Colbert}, S.~C., \& {Varoquaux}, G. 2011, Computing in
  Science Engineering, 13, 22

\bibitem[{Van~Rossum \& Drake(2009)}]{python}
Van~Rossum, G. \& Drake, F.~L. 2009, Python 3 Reference Manual (Scotts Valley,
  CA: CreateSpace)

\bibitem[{{Virtanen} {et~al.}(2020){Virtanen}, {Gommers}, {Oliphant},
  {Haberland}, {Reddy}, {Cournapeau}, {Burovski}, {Peterson}, {Weckesser},
  {Bright}, {van der Walt}, {Brett}, {Wilson}, {Jarrod Millman}, {Mayorov},
  {Nelson}, {Jones}, {Kern}, {Larson}, {Carey}, {Polat}, {Feng}, {Moore}, {Vand
  erPlas}, {Laxalde}, {Perktold}, {Cimrman}, {Henriksen}, {Quintero}, {Harris},
  {Archibald}, {Ribeiro}, {Pedregosa}, {van Mulbregt}, \&
  {Contributors}}]{scipy}
{Virtanen}, P., {Gommers}, R., {Oliphant}, T.~E., {et~al.} 2020, Nature
  Methods, 17, 261

\bibitem[{{Vlahakis} \& {K{\"o}nigl}(2004)}]{Vlahakis2004}
{Vlahakis}, N. \& {K{\"o}nigl}, A. 2004, \apj, 605, 656

\bibitem[{{Woo} \& {Urry}(2002)}]{Woo2002}
{Woo}, J.-H. \& {Urry}, C.~M. 2002, \apj, 579, 530

\end{thebibliography}

\begin{table*}
\caption{Model-fit parameters for the 86 GHz GMVA observations.}             
\label{table1}      
\centering                          
\begin{tabular}{c c c c c c c c}        
\hline\hline                 
Epoch & Flux & $\delta$F & Distance & $\delta$D & PA & FWHM & ID \\
(year) & (Jy) & (Jy) & (mas) & (mas) & (deg) & (mas) \\
 2009.35 & 0.790 & 0.079 & -- & -- & -- & 0.003 & C0 \\
 & 0.262 & 0.027 & 0.066 & 0.005 & -168.18 & 0.014 & d1 \\
 & 0.273 & 0.028 & 0.169 & 0.006 & -168.34 & 0.032 & d2 \\
 & 0.043 & 0.004 & 0.270 & 0.005 & -172.65 & 7.3e-8 & d3 \\
 & 0.390 & 0.040 & 0.339 & 0.009 & -164.98 & 0.077 & d4 \\
 \hline
 2009.77 & 1.417 & 0.142 & -- & -- & -- & 0.019 & C0 \\
 & 0.427 & 0.043 & 0.176 & 0.007 & -176.05 & 0.058 & d2 \\
 & 0.141 & 0.016 & 0.351 & 0.008 & -171.52 & 0.039 & d4 \\
 \hline
 2010.35 & 1.098 & 0.110 & -- & -- & -- & 0.023 & C0 \\
 & 0.411 & 0.042 & 0.115 & 0.007 & -174.26 & 0.052 & d1 \\
 & 0.392 & 0.040 & 0.292 & 0.019 & -173.50 & 0.172 & d3 \\
 \hline
 2011.35 & 1.790 & 0.179 & -- & -- & -- & 0.027 & C0 \\
 & 0.447 & 0.045 & 0.186 & 0.006 & -172.41 & 0.050 & d2 \\
 & 0.381 & 0.039 & 0.385 & 0.009 & -172.09 & 0.075 & d4 \\
\hline
 2012.38 & 1.421 & 0.142 & -- & -- & -- & 0.066 & C0 \\
 & 0.320 & 0.033 & 0.200 & 0.015 & -165.73 & 0.128 & d2 \\
 \hline
 2013.74 & 0.747 & 0.075 & -- & -- & -- & 0.018 & C0 \\
 & 0.425 & 0.043 & 0.148 & 0.007 & 171.89 & 0.052 & d1 \\
 & 0.490 & 0.050 & 0.304 & 0.008 & 175.13 & 0.072 & d3 \\
 \hline
 2015.37 & 1.001 & 0.100 & -- & -- & -- & 0.032 & C0 \\
 & 0.332 & 0.034 & 0.072 & 0.007 & -157.54 & 0.050 & d1 \\
 & 0.194 & 0.021 & 0.196 & 0.009 & -170.31 & 0.054 & d2 \\
 & 0.287 & 0.030 & 0.379 & 0.013 & -169.43 & 0.105 & d4 \\
 \hline
 2015.73 & 0.488 & -- & -- & -- & -- & 0.019 & C0 \\
 & 0.188 & 0.020 & 0.116 & 0.009 & 175.71 & 0.059 & d1 \\
 & 0.116 & 0.015 & 0.303 & 0.019 & -166.19 & 0.091 & d3 \\
 \hline
 2016.39 & 0.461 & 0.046 & -- & -- & -- & 3.0e-8 & C0 \\
 & 0.078 & 0.012 & 0.190 & 0.022 & -173.84 & 0.086 & d2 \\
 & 0.067 & 0.012 & 0.377 & 0.037 & -167.58 & 0.124 & d4 \\
 & 0.020 & -- & 0.124 & 0.005 & -5.70 & 2.7e-8 & u1 \\
 \hline
 2016.75 & 0.525 & 0.053 & -- & -- & -- & 0.041 & C0 \\
 & 0.227 & 0.024 & 0.148 & 0.020 & -165.71 & 0.136 & d1 \\
 \hline
 2017.25 & 0.706 & 0.071 & -- & -- & -- & 0.027 & C0 \\
 & 0.470 & 0.048 & 0.123 & 0.008 & -160.43 & 0.072 & d1 \\
 & 0.078 & 0.012 & 0.213 & 0.015 & 155.43 & 0.062 & d2 \\
 & 0.287 & 0.030 & 0.312 & 0.010 & -165.82 & 0.078 & d3 \\
 & 0.075 & -- & 0.114 & 0.010 & -19.17 & 0.040 & u1 \\
 \hline
 \end{tabular}
\end{table*}



\appendix
\section{The jet collimation profile in BL Lac using AIPS} \label{app:A}
\begin{figure}
 \centering
  \includegraphics[width=0.5\textwidth]{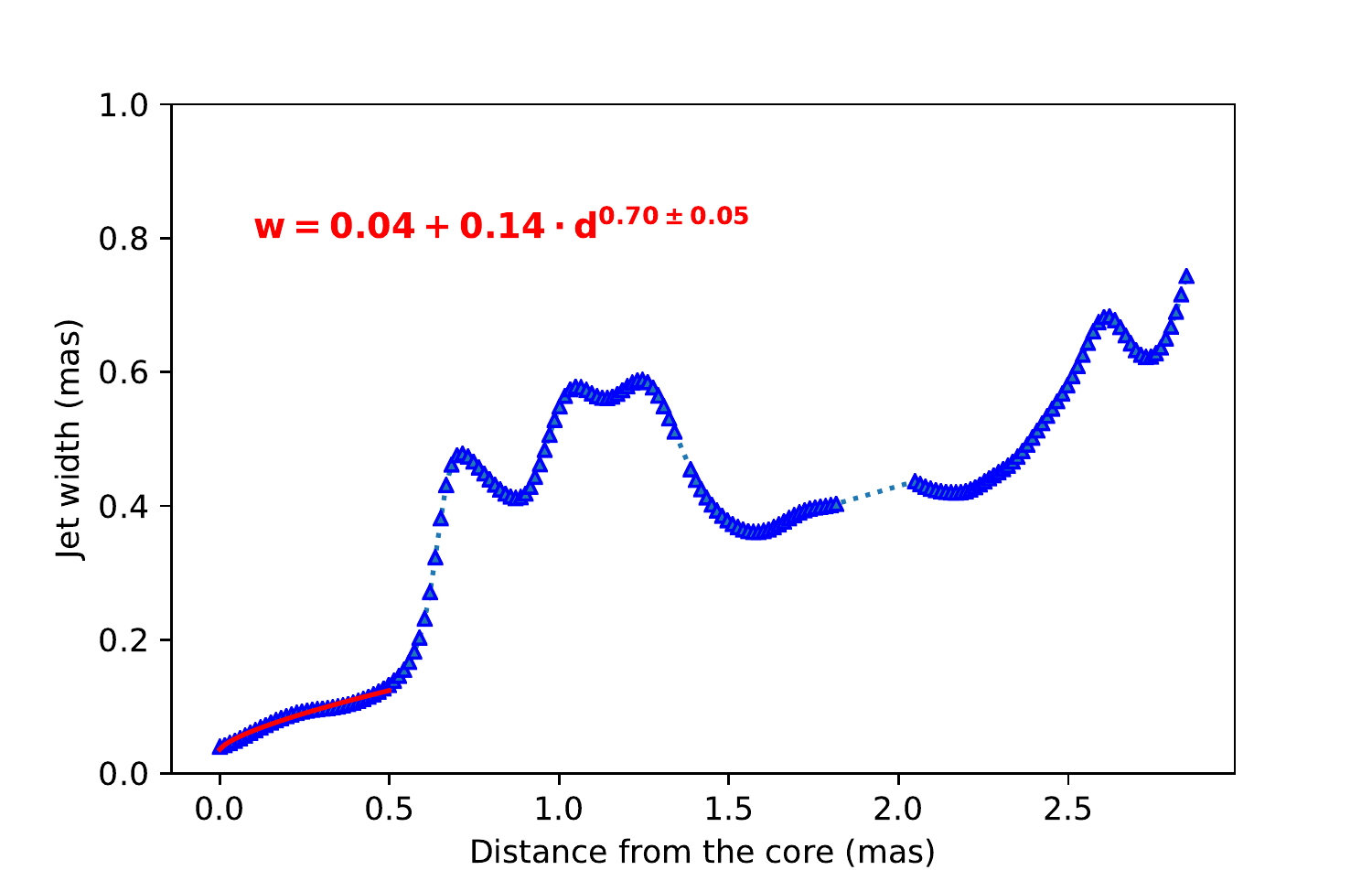}
  \caption{Jet width versus distance along the jet in the 43 GHz VLBA stacked image of BL Lac, obtained using AIPS.}
  \label{43jetprof_aips}%
\end{figure}

\begin{figure}
 \centering
  \includegraphics[width=0.5\textwidth]{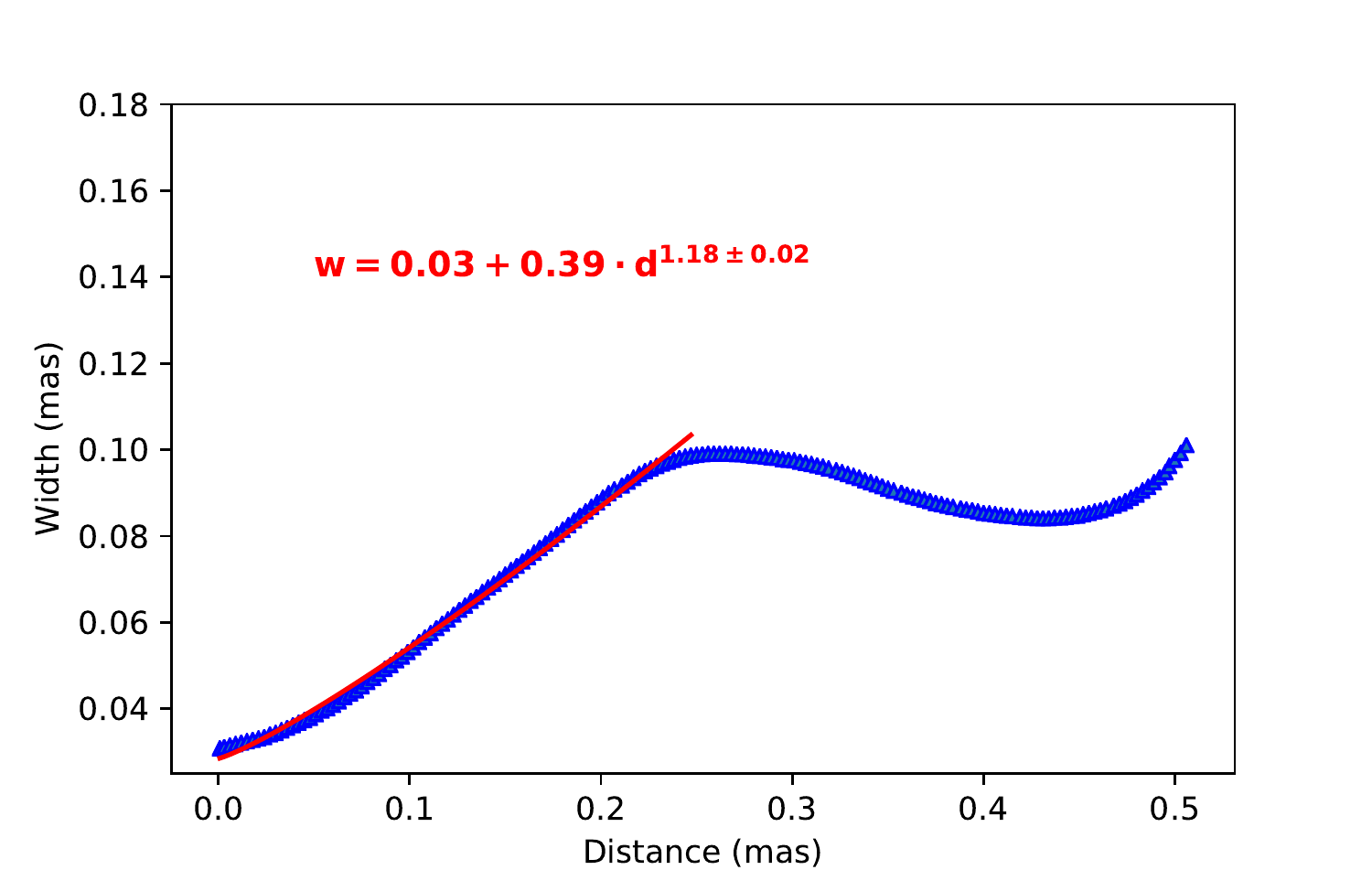}
  \caption{Jet width versus distance along the jet in the 86 GHz VLBA stacked image of BL Lac, obtained using AIPS.}
  \label{86jetprof_aips}%
\end{figure}

Figures~\ref{43jetprof_aips} and \ref{86jetprof_aips} show the jet collimation profiles at 43 and 86 GHz, respectively, obtained using the tasks SLICE and SLFIT in AIPS. Both profiles as well as the fit parameters, obtained using the least squares method, are in agreement with what obtained using the ridge line method (see Section~\ref{jetprofile} and Figures~\ref{43jetprof} and \ref{86jetprof}). 
The difference here is that we could not fit the \textit{w-d} dependence after 1.5 mas in the 43 GHz profile because of some missing values around 2 mas, where AIPS was not able to fit properly the brightness profile associated with some slices. 

\end{document}